\documentclass[twocolumn,UKenglish,texlive=2011,txfonts,11pt]{article}
\pdfoutput=1

\usepackage{defs}
\usepackage{graphicx}
\graphicspath{{./figures/}{../figures/}}

\title{Citizen Scientist Community Engagement with the 
HiggsHunters project at the Large Hadron Collider}

\begin{document}
\author[a]{A.J~Barr}
\author[b]{A.~Haas}
\author[a,c]{C.W.~Kalderon}
\affil[a]{Department of Physics, University of Oxford, Oxford, UK}
\affil[b]{Department of Physics, NYU, New York, USA}
\affil[c]{Department of Physics, University of Lund, Sweden}

\maketitle

\abstract{%
The engagement of Citizen Scientists
with the \url{HiggsHunters.org} citizen science project
is investigated through analysis of behaviour,
discussion, and survey data. 
More than \nvolunteers{} Citizen Scientists from \ncountries{} countries
participated, classifying \nclicks{} features of interest
on about \nimagesviewed{} distinct images.
While most Citizen Scientists classified only a handful 
of images, some classified hundreds or even thousands.
Analysis of frequently-used terms on the dedicated
discussion forum demonstrated that 
a high level of scientific engagement was not uncommon.
Evidence was found for a emergent and distinct technical vocabulary
developing within the Citizen Science community.
A survey indicates a high level of engagement and an appetite 
for further LHC-related citizen science projects.
}

\section{Introduction}
\label{sec:intro}

The Large Hadron Collider is arguably the highest profile scientific project of our time.
The discovery of the Higgs boson~\cite{HIGG-2012-27,CMS-HIG-12-028} has been the scientific highlight to date.
The accelerator continues to be the subject of much media attention as searches for other new particles continue.

\begin{figure}
\centering
\includegraphics[width=0.98\linewidth]{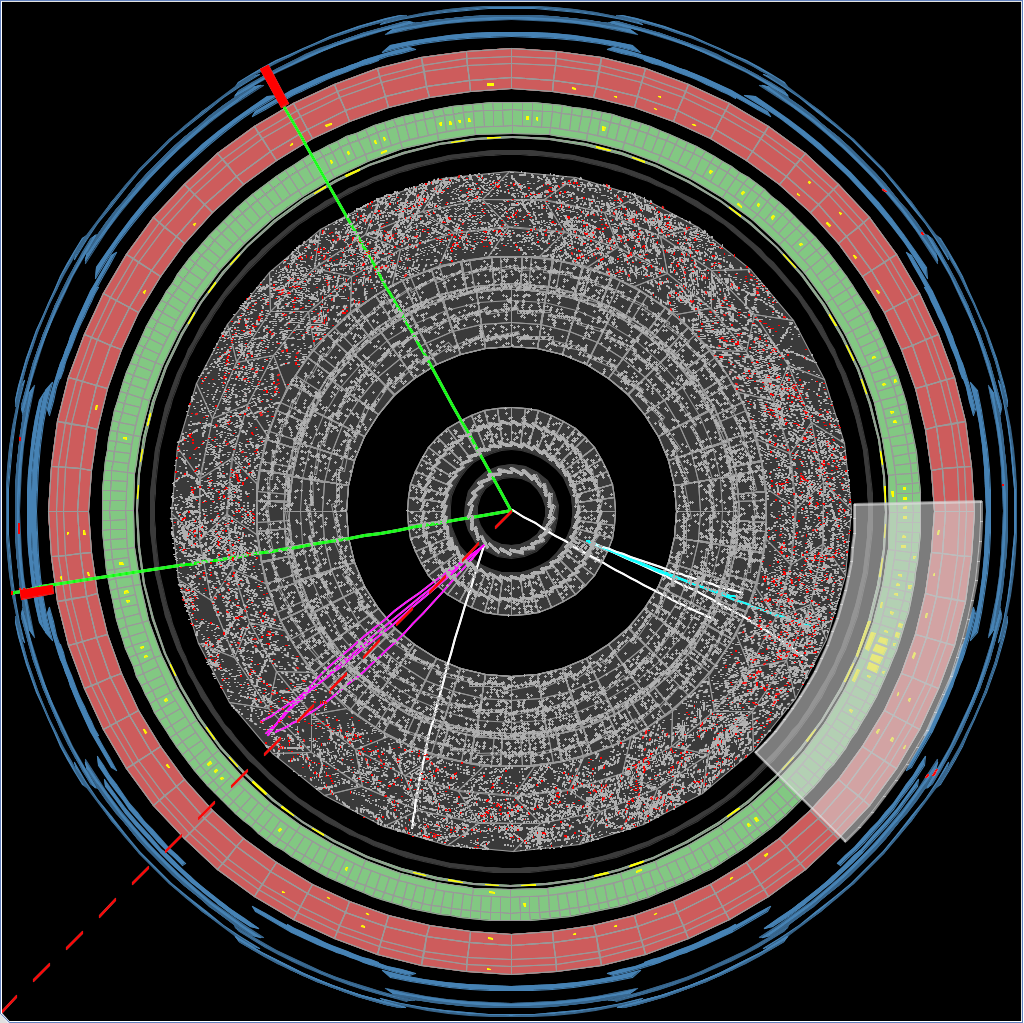}
\caption{\label{fig:exampleimage}
An example ATLAS detector image presented to citizen scientists. 
This image contains two off-centre vertices, 
each visible as a vee-like structure, 
at about 4 o'clock and 7 o'clock, a little distance from the center of the image.
The image was generated from a computer simulation.
}
\end{figure}

Matching this cutting-edge science with the public's curiosity to understand it can present a challenge. 
The particles created at the LHC are themselves invisible. 
Many, including the Higgs boson, 
decay a tiny fraction of a second after their creation, and can only be detected and reconstructed 
using large dedicated detectors assembled over decades by large international collaborations.

Nevertheless, there is a strong drive within science policy to 
allow the public to be involved in not just reading about science,
but actually performing it.
Citizen science projects --
which directly involve the public in the scientific process --
represent an ideal vehicle for meaningful engagement with a large community.
Particular citizen science projects previously have been shown to 
reveal that participants were engaged in thinking processes
similar to those of scientific investigations~\cite{Trumbull:2000}.
Crowdsourced research has itself been shown to be reliable, scalable, 
and connective~\cite{Watson2016}.

When considering what might be viable citizen science
projects for the particular case of the LHC reported here, 
several factors were considered.
The subject matter should be sufficiently appealing to 
attract a sufficient number of citizen scientists.
The tasks assigned to the citizen scientists must
be within their capability, or possible to be 
rapidly understood, to maintain volunteers' interest.
And to motivate continued engagement there
should be the possibility of making a very significant
contribution to knowledge.

It was noted that citizen scientists have previously been shown 
to be good classifiers of images~\cite{zooniverse_papers}. 
They are also efficient at spotting unusual objects in images
including unexpected galaxy features~\cite{Lintott11102009}.
Through the Galaxy Zoo~\cite{galaxyzoo} project alone, citizen scientists 
have contributed to the results of 48 scientific 
papers~\cite{zooniverse_papers}.
The present study evaluates, using the data from the \url{HiggsHunters.org} project 
described below,
the extent to which analysis by 
citizen scientists might also be possible at the Large Hadron Collider,
and the engagement of those citizen scientists with that subject matter.

Previously within the field of particle physics, 
the public has been invited to contribute to 
CERN's science by donating idle time on their computer to help simulate 
proton-proton collisions~\cite{lhcathome,atlasathome}.
That project aids the scientific endeavour, however the
the volunteers are providers 
of computing resource rather than active researchers.
More direct involvement in the research has previously been
restricted to the relatively small fraction of the public that has a high level of 
computing coding skills.
Such individuals have been able to directly analyse data from CERN experiments
via the CERN opendata portal~\cite{opendata}.
The Kaggle project~\cite{kaggle} in which members of the public were challenged 
to use machine learning to identify Higgs boson events
was very successful, but also demanded a high level of coding expertise,
making it inaccessible to most members of the public.

The HiggsHunters project is, to the best of our knowledge, the first to allow the 
non-expert general public a direct role in searching for new particles at the 
LHC.


\begin{figure}
\centering\begin{minipage}{0.9\linewidth}
\begin{framed}
\begin{center}\textbf{`Baby' bosons}\end{center}
\vskip 2mm
The physics theories under test
predict the existence of hypothetical new particles $\phi$ which are not in the 
Standard Model of particle physics and which 
have not yet been observed experimentally.
In such theories the usual Higgs boson $H$, after it is created, 
would most often decay as predicted by the Standard Model,
however a fraction of the time it would decay into 
the new particles:
\[
H \to \phi + \phi.
\]
The new particles $\phi$ interact with the Standard Model only very weakly.
This weak coupling means they have a slow decay rate,
and hence a relatively long lifetime on the particle scale
-- typically of order nanoseconds.
They can therefore travel a macroscopic distance,
perhaps tens of centimetres, before themselves decaying.
\end{framed}
\end{minipage}
\end{figure}

For the \url{HiggsHunters.org} project, 
a task was created which lent itself well to the strengths
of non-expert citizen scientists -- in particular their abilities
to classify elements in images, and to spot unusual features.

The task selected
was to ask citizen scientists to identify any sets of tracks originating from 
points away from the centre of the image -- known as 
{\em Off-Centre Vertices} (abbreviated OCV).
Such tracks can be observed in 
the image of a simulated collision shown in \cref{fig:exampleimage}.
Such features indicate the presence of a relatively long-lived neutral particle,
which travelled some centimetres from the interaction point at the centre of the image
before decaying producing spray of a large number of tracks.

Collective evidence from the body of citizen scientists 
about these OCVs could indicate
new particles beyond the knowledge of particle physics --
dramatically changing our understanding of the subatomic realm.
The high impact of a potential discovery 
meets the important motivating feature of citizen science projects that 
the volunteers have a real opportunity of discovering something previously 
unknown to science~\cite{Cox:2015}.
It also satisfies the ethical criterion~\cite{Riesch:2014}
that the time of the citizen scientists is being used productively.

The citizen scientists were also given the task of identifying
anything they thought was `weird' in any image.
Serendipity can have an important role in scientific discovery,
so it was considered important to flag such particularly unusual features.

The citizen science web interface 
was constructed within the Zooniverse~\cite{Simpson:2014:ZOW:2567948.2579215}
framework, using images from the ATLAS experiment at the Large Hadron Collider.
Both images from real collisions and those from Monte Carlo simulations
were displayed, with the Citizen Scientist being unaware (at the time of classification)
as to whether the image was based on real or simulated data.
The ability of volunteers to identify the off-centre vertices
could then be calibrated using the test images which showed simulations
of the decay processes of interest. 

All images, whether simulation or from real collisions, were processed 
using the ATLAS reconstruction software~\cite{SOFT-2010-01}, 
with some additions~\cite{Aad:2015rba}.

\section{Citizen Scientist behaviour}
\label{sec:analysis:behaviour}

As of October 2017, classifications had been performed by 57,613 Citizen Scientists, 
of whom 25,608 had created Zooniverse accounts.
New Citizen Scientists 
are invited to create a Zooniverse account after their first five classifications, 
and periodically thereafter.
For those classifications made without Zooniverse accounts 
it is assumed that classifications from different IP 
addresses are distinct scientists.

\begin{figure}
\centering
\begin{subfigure}[b]{0.49\textwidth}\centering
\includegraphics[width=0.98\textwidth]{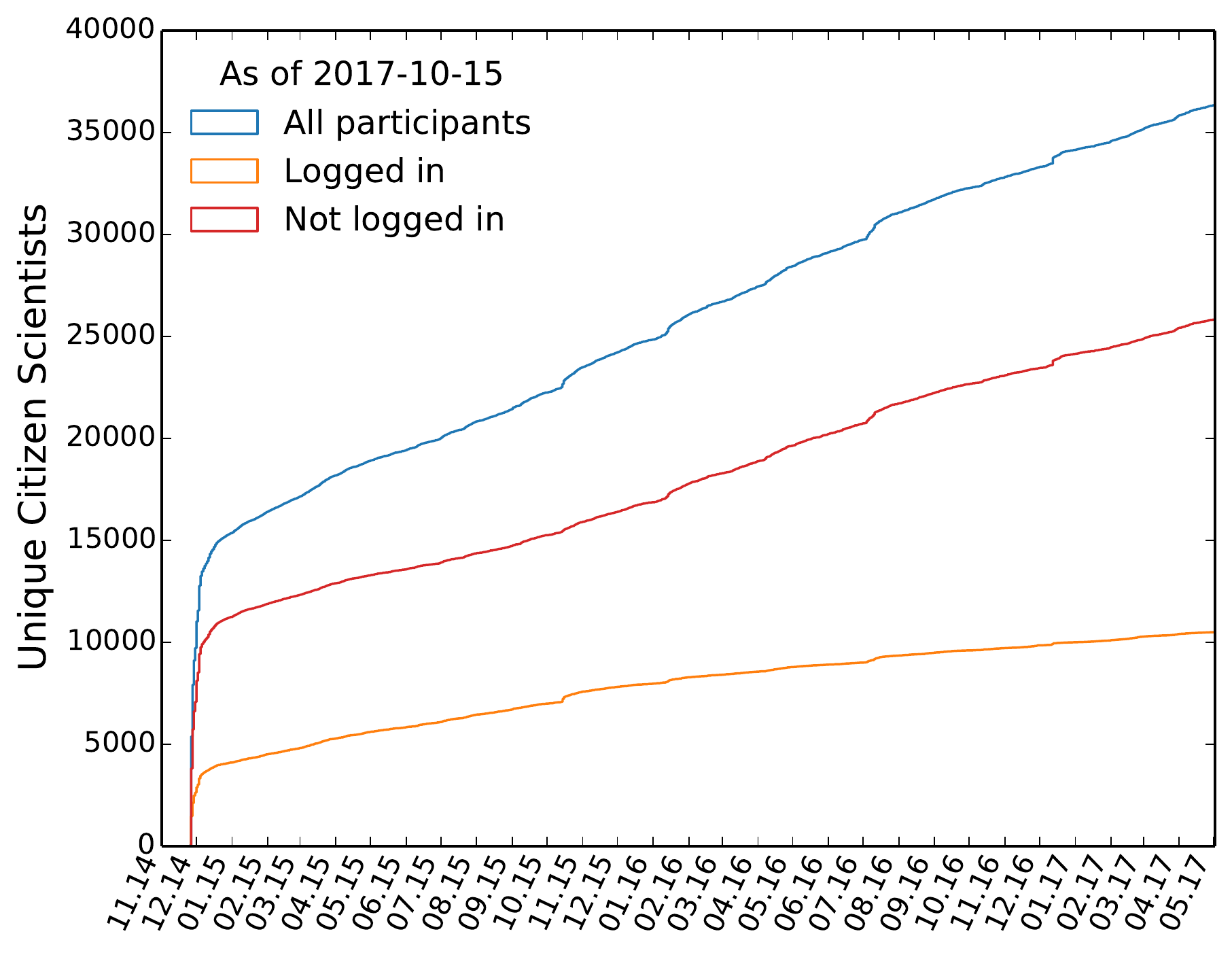}
\end{subfigure}
\begin{subfigure}[b]{0.49\textwidth}\centering
\includegraphics[width=0.98\textwidth]{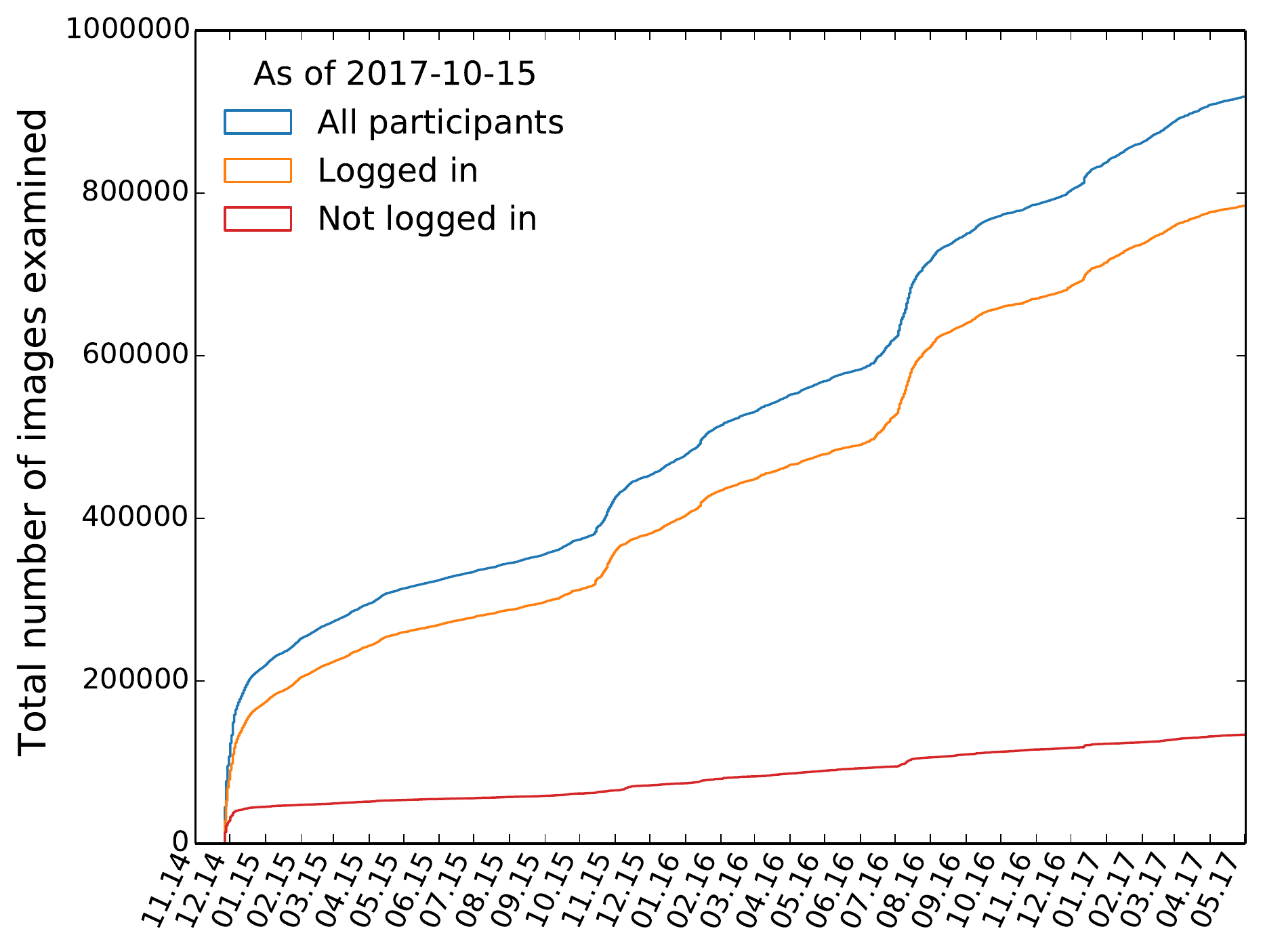}
\end{subfigure}
\caption{\label{fig:timeline}
(a) Cumulative number of unique Citizen Scientists as a function of time.
(b) Cumulative number of images examined as a function of time.
}
\end{figure}

\begin{figure}
\centering
\begin{subfigure}[b]{0.49\textwidth}\centering
\includegraphics[width=0.98\textwidth]{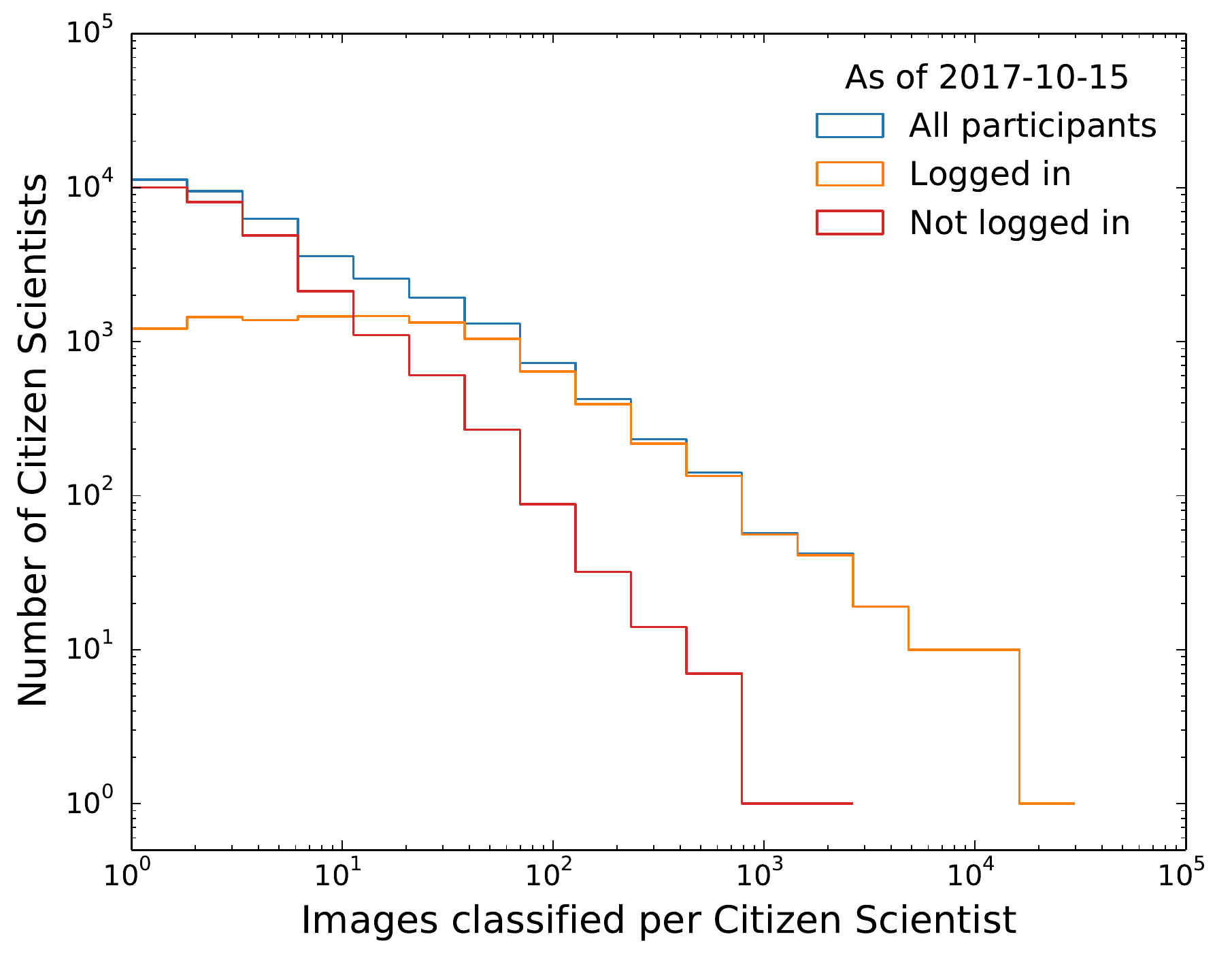}
\end{subfigure}
\caption{\label{fig:clicks_per_user}
Number of classifications per Citizen Scientist.
}
\end{figure}

The number of Citizen Scientists is shown in 
\cref{fig:timeline}, as is the cumulative number of classifications.
There was a rapid rise in the number of new scientists soon after the project launch,
when the project was advertised via email to existing Zooniverse account holders as well
as in the press and social media.
Subsequent periods during which new Citizen Scientists are attracted are observed, 
for example in July 2016 when a CERN news story
was published about the project~\cite{cern-article}.

The number of classifications per Citizen Scientist (\cref{fig:clicks_per_user})
follows an approximate power-law behaviour. 
Most volunteers dip in to classify just a handful of images,
though more than a thousand individuals provided one hundred or 
more classifications.
At the upper end of the distribution, 
more than one hundred volunteers provided more than 1,000 classifications, 
with the most dedicated enthusiast providing more than 25,000 classifications.

Several moderators were selected from among the 
Citizen Scientists active on a dedicated `Talk' 
discussion forum~\cite{HiggsHuntersTalk} 
to help answer questions from other, less experienced volunteers.
The moderators helped newer volunteers with identification of objects,
and with some of their science questions.
Other scientific questions were addressed by the science team, either 
via the Talk forum or in the project's blog forum~\cite{blog}.

\section{Science Objectives}
\label{sec:analysis:ability}

An initial determination~\cite{Barr:2016vce}
has previously been made of the performance of citizen scientists 
relative to computer algorithms that were developed and used by the ATLAS 
collaboration to identify off-centre vertices~\cite{Aad:2015rba}. 

It was found that the performance of the Citizen Scientists competed
very well with that of the computer algorithm.
The collective ability of the Citizen Scientists 
was superior to the ATLAS computer algorithm 
for simulations with low-mass long-lived particles.
A detailed comparison of the identification performance of the Citizen Scientists
relative to the computer algorithm is described in Ref~\cite{Barr:2016vce}.

In addition to being able to mark off-centre vertices, 
the Citizen Scientists are also encouraged to select anything `weird' in the images, 
and to follow up these on the Talk forum
where the wider community discusses them. 
This raised several instances of known phenomena, 
such as cosmic ray showers passing through ATLAS, 
but also some that were unexpected, demonstrating the potential 
for untrained Citizen Scientists to isolate interesting features in real LHC collision data.




\section{Citizen Scientist Discussion}
\label{sec:analysis:comments}
The Zooniverse platform provides a forum for Citizen Scientists to 
build community, discuss objects and images, and to ask questions.
The forum is open to all Citizen Scientists, moderators and 
project scientists.

An analysis was performed of the content of the 20,257 comments
received between November 2014 and May 2017.
These comments were received from 1345 different Citizen Scientists.
The distribution of the number of words per comment is found to
follow a falling exponential form with a mean of 6.6 words, 
and with 6\% of comments being 20 words or more,
which suggest substantial observations and/or questions.

\begin{figure}
\begin{subfigure}[b]{0.49\textwidth}\centering
\includegraphics[width=0.98\textwidth]{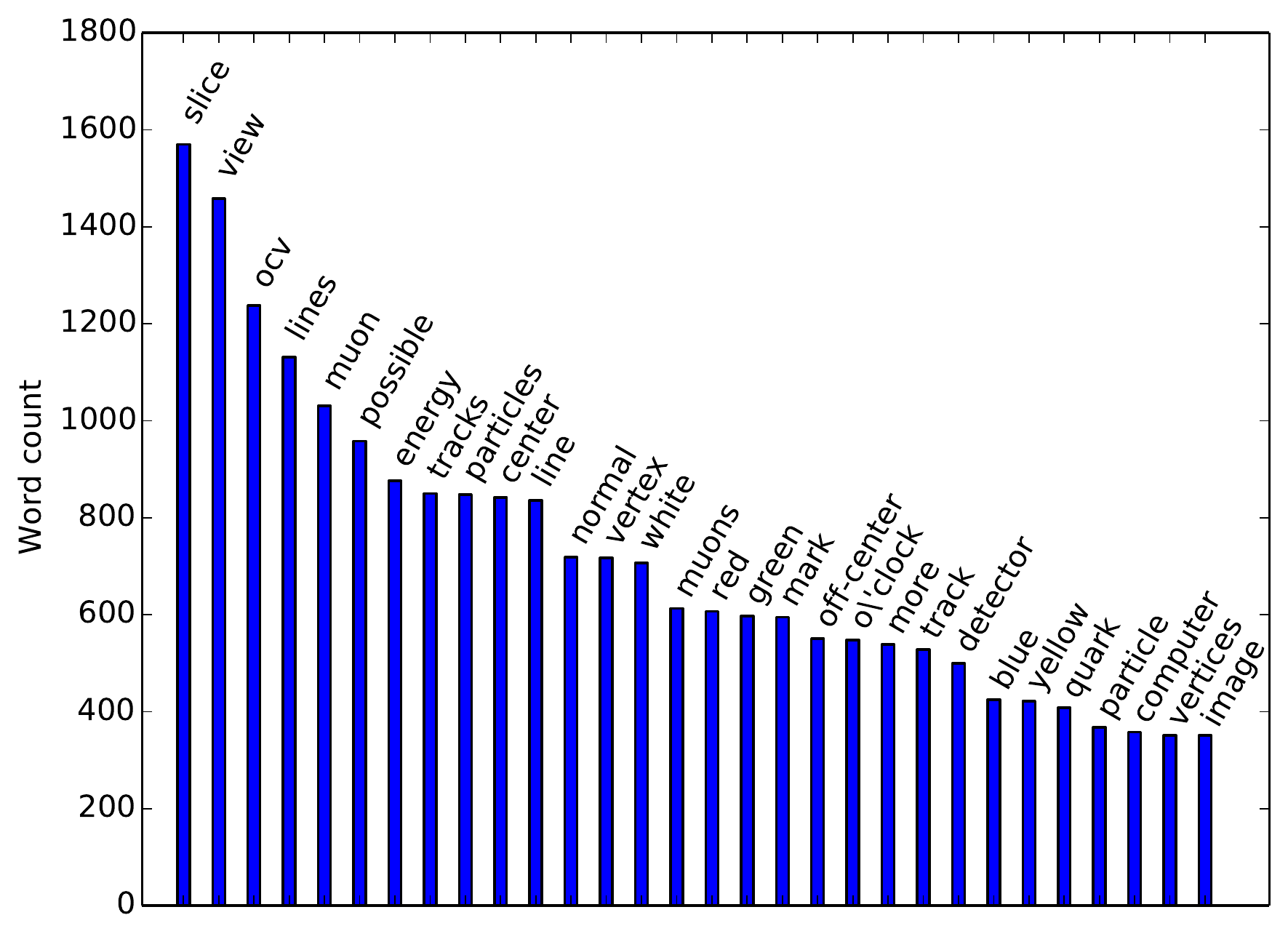}
\end{subfigure}

\begin{subfigure}[b]{0.49\textwidth}\centering
\includegraphics[width=0.98\textwidth]{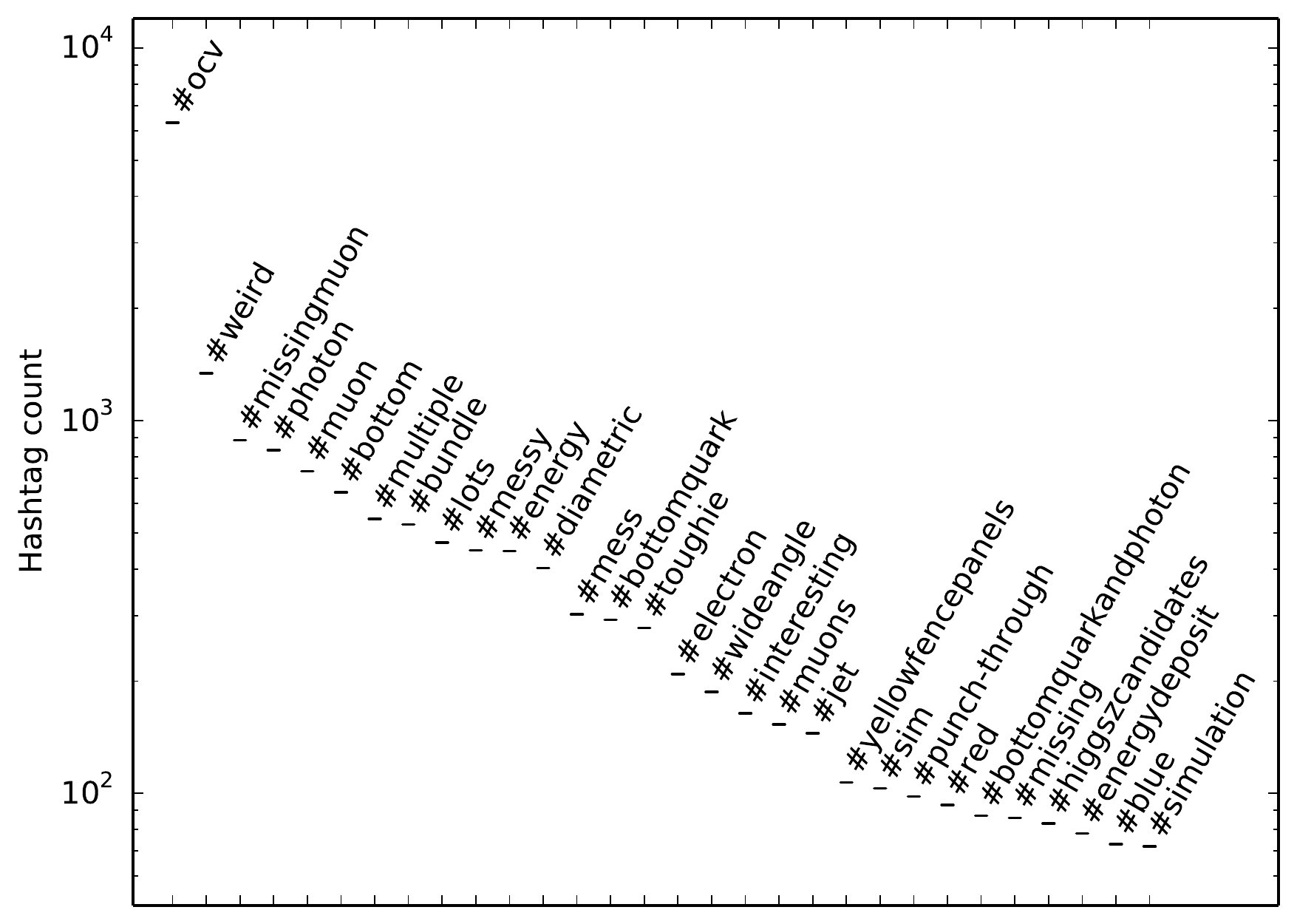}
\end{subfigure}
\caption{\label{table:frequencies}
Frequently used words (above) and hashtags (below),
and their frequency of use, in 20,257 Talk forum comments.
The most common words used in everyday language such as `a' `the', and `of' are omitted.
}
\end{figure}

Frequently used words and hashtags are shown in 
\cref{table:frequencies} 
The most common hashtags are `\#ocv' and `\#weird', which indicate
the two proposed tasks of identifying off-centre-vertices and  unusual features
respectively.

Other common words and hashtags include those denoting basic image features 
and descriptors such as `white' `line' and `image'. 
Others describe the Citizen Scientists' impressions of 
or reactions to the images such as `\#toughie', `\#mess' and `\#interesting'.
A high level of insight and learning is demonstrated
through the use of more technical and abstract terms
such as `\#tracks', `\#muons' `\#electrons' and `\#energy', which are
physics objects represented in the images.
The hashtags `\#higgszcandidates', `\#punchthrough',
and `\#bottomquarkandphoton'
are highly technical and suggest a level of understanding 
similar to that of a particle physics professional.

The meanings of some terms used frequently by Citizen Scientists were later formalised
by a Citizen Scientist moderator in a Talk post, including:
\begin{quote}
{\bf \#bundle}: Several particle tracks that appear to share a common origin, but do not meet at a vertex.

{\bf \#diametric}: Many particles (or lots of energy) located on opposite sides of the detector, with relatively little between.

{\bf \#messy}: Objects which are complicated by many crossing lines, which can make it difficult to find off-centre vertices.
\end{quote}

The hashtag \#diametric, used by 29 Citizen Scientists in 442 comments, 
was adopted by Citizen Scientists 
to describe what in the physics literature is called a `two-jet event'.
The term `\#bundle' was used by 43 different Citizen Scientists in 619 different comments.
Unlike `diametric', the term `bundle' is also used 
in a technical sense in the general particle physics literature,
but in a slightly different context --- to indicate
sets of near-collinear tracks but in the context of cosmic ray showers,
rather than collider physics.

\begin{figure}
\centering
\begin{subfigure}[b]{0.49\textwidth}\centering
\includegraphics[width=0.98\textwidth]{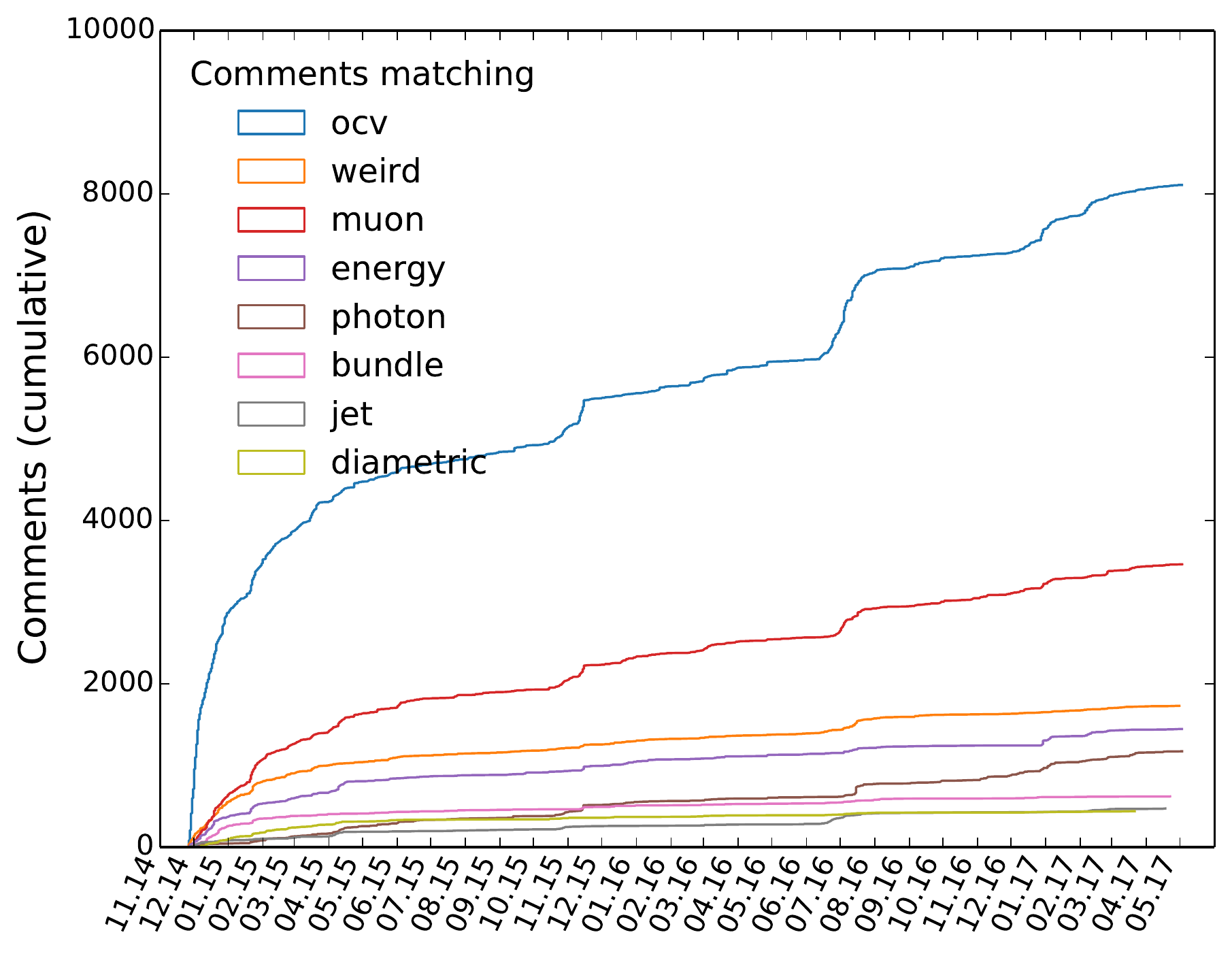}
\end{subfigure}
\caption{\label{fig:comments}
Cumulative number of comments matching particular words as a function of date.
}
\end{figure}

\begin{figure}
\centering
\begin{subfigure}[b]{0.49\textwidth}\centering
\includegraphics[width=0.98\textwidth]{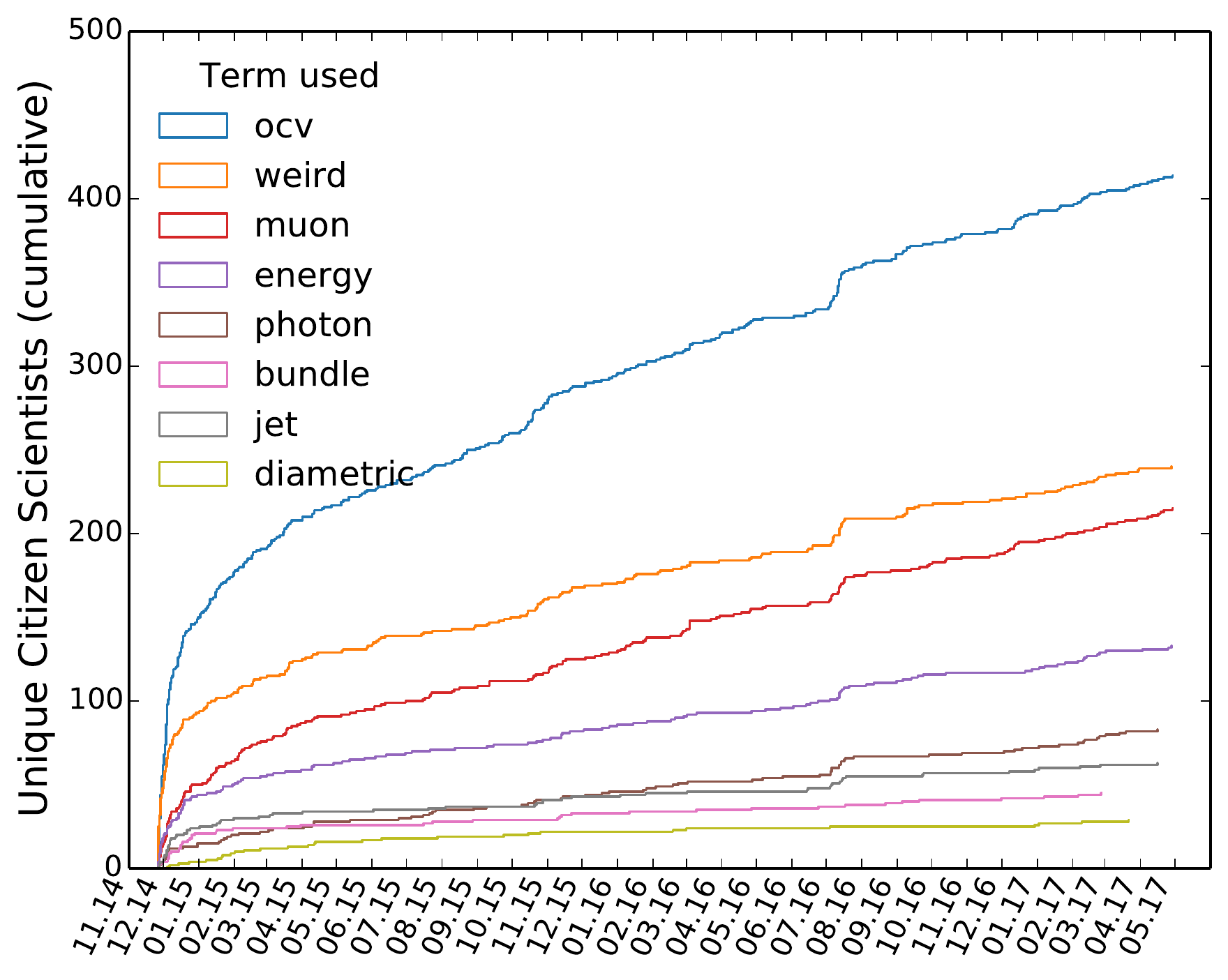}
\end{subfigure}
\caption{\label{fig:commenters}
Cumulative number over time of unique Citizen Scientists using particular words.
}
\end{figure}

The cumulative distribution of the use of particular terms over 
time~(\cref{fig:comments})
shows different types of use at different times. 
For example the cumulative frequency of use of the term ``weird'' has a tendency to flatten out
with time (presumably as Citizen Scientists become accustomed to particular features)
whereas the cumulative counts of some 
technical terms such as ``photon'' and ``muon'' keep growing rapidly.
The number of unique Citizen Scientists using particular words has continued
to grow with time (\cref{fig:commenters}). 
Seemingly the non-standard term ``bundle''
fell out of fashion after the first couple of months, 
being overtaken by the term ``jet'' which is the usual word for this
feature within the wider particle physics community.

\section{Survey Evaluation}
\label{sec:analysis:evaluation}

To evaluate the impact of the project on the Citizen Scientists themselves, 
a web-based survey was was undertaken,  
with an invitation to participate being sent to all registered HiggsHunters volunteers. 
The number of respondents was 322 (including 63 partial responses). This response rate represents about 1\% of those who participated as Citizen Scientists in the project. The survey was advertised via the Zooniverse website and in an email to those with Zooniverse accounts, which is likely to have led to some bias towards respondents having a higher degree of engagement than average. This supposition is supported by the observation that about 80\% of survey respondents had previously participated in another Zooniverse project prior to HiggsHunters. 

The gender of respondents was 33\% female and 65\% male (with 2\% preferring not to say). 
A wide range of ages was represented (\cref{tab:ages}).
This is also reflected in the diversity of occupations, with  
19\% of respondents being students, 37\% in full-time work, and 22\% of respondents retired
(with the remainder having other employment status).
Well-represented occupations included teachers, engineers, consultants, developers and researchers.
Respondents tended to be well educated: 74\% have at least 
an undergraduate degree,
39\% had at least a masters degree and 14\% held a doctoral degree. 
It was notable that only about a quarter of those holding a masters
degree or higher held that degree in a physics-related subject,
showing that the project had appeal to those trained in other disciplines, 
particularly in other areas of science, technology, engineering and mathematics.

The best-represented countries were the USA (25\%) and the UK (16\%),
with a total of 35 countries represented amongst all respondents. 
A bias towards native English speakers (65\% of respondents) was 
perhaps unsurprising given that the \url{HiggsHunters.org} website is only 
available in the English language.

\begin{table}
\centering
\begin{tabular}{lll}
Age 	& Percent & 	Count \\
\hline
16 to 17 &	7\% 	&	17 \\ 
18 to 19 &	3\% 	&	8 \\ 
20 to 24 &	5\% 	&	14 \\ 
25 to 34 &	15\% 	&	39 \\ 
35 to 44 &	15\% 	&	37 \\ 
45 to 54 &	15\% 	&	38 \\ 
55 to 64 &	21\% 	&	53 \\ 
65 to 74 &	12\% 	& 30 \\
75 or older & 	4\% 	& 11 \\
Prefer not to say & 	3\% &	7 \\ 
\end{tabular}
\caption{\label{tab:ages}
Age distribution of survey respondents.
}
\end{table}



Of the respondents, 80\% had engaged in citizen science before, in another science area, 
while for 20\% it was their first citizen science project.  
About 62\% were native English speakers, but many other native languages were represented. 
Geographically, 33 were based in the USA, 21\% in the UK, and many other countries were also represented.

\begin{table}
\centering\begin{tabular}{lll}
Change &	Percent &	Count \\
\hline
A lot           & 13\% 	& 35 \\
Moderately 	& 14\% 	& 37 \\
Slightly 	& 20\% 	& 54 \\
No change 	& 36\% 	& 98 \\ 
N/A 	& 17\% 	& 46 \\
\end{tabular}
\caption{\label{tab:study}
Responses to the question ``To what extent are you more likely to study physics in the future as a result of participating in HiggsHunters?''. 
}
\end{table}

\begin{table}
\centering\begin{tabular}{lll}
Discussed\ldots &	Percent & Count \\ 
\hline
\ldots with your family? & 58\% &	99 \\
\ldots with friends? & 64\% &	111 \\
\ldots with colleagues? & 32\% & 56\\ 
\ldots on social media? & 15\% & 26 \\
\end{tabular}
\caption{\label{tab:discussed}
Answers of survey respondents to the question ``Have you ever discussed Zooniverse\ldots''
} 
\end{table}

More than 80\% of respondents indicated that their knowledge of particle physics had been 
improved to some extent as a direct result of participating in HiggsHunters. 
In terms of future directions, 47\% of respondents said they were more likely (to some extent) 
to go on to study physics as a result of participating in the project (\cref{tab:study}).
This can be considered a high fraction, given the broad age range of participants.

In terms of dissemination, many respondents had discussed the project with others, 
including friends, family and work colleagues (\cref{tab:discussed}). This indicates
the project had a multiplier effect, in that it reached
more people than just those citizen scientists directly involved.
This willingness to discuss with others also indicates a high level of 
feeling of ownership 
and interest among the citizen scientists themselves.

The expectation that the survey respondents were subject to a selection bias
(compared to the general population of HiggsHunters citizen scientists)
towards more highly engaged end of the spectrum is confirmed from their responses to a question
asking about the duration of the period during which they performed classification
(\cref{tab:duration}). That distribution for respondents is more broadly distributed than 
would be expected from the general population of Citizen Scientists, 
which peaks at low numbers of classifications (\cref{fig:clicks_per_user}).
Nevertheless, the wide range of different levels of duration
among the respondents show that an interesting section of the Citizen Scientists has been
being sampled, albeit with some bias.
No attempt has been made to extrapolate to the general population,
since with the numbers of people surveyed, insufficient information is available about
possible confounding factors which could significantly affect
that extrapolation.

\begin{table}
\centering\begin{tabular}{lll}
Duration &	Percent &	Count\\ 
\hline
A single session &		10\% 	& 26 \\
One or two days &		14\% 	& 35 \\
2-7 days &		15\% 	& 39 \\
2-4 weeks &		17\% 	& 44 \\
1-5 months &	 	18\% 	& 45 \\
6-12 months &	 	10\% 	& 26 \\
Over a year &	 	15\% 	& 37 \\
\end{tabular}
\caption{\label{tab:duration}
Response to the question ``Over what duration did you classify images?''
}
\end{table}

\begin{table*}
\centering\begin{tabular}{lll}
Subsequent activity &	Percent &	Count\\ 
\hline
Read or watched more about science & 87\% & 	152 \\
Studied science more formally 	& 29\% &	51 \\
Carried out your own research 	& 20\% &	35 \\
Attended lectures or similar events &	19\% 	& 33 \\
Attended science fairs or similar events & 15\% & 26 \\
\end{tabular}
\caption{\label{tab:future}
Response to the question ``{\em As a result of the HiggsHunters project}, have you done any of the following?''
}
\end{table*}

\begin{table}
\centering\begin{tabular}{lll}
Subsequent projects & Percent & Count \\
\hline
None (at time of response) & 22\%  & 62 \\ 
Zooniverse project(s) & 74\% & 209 \\
non-Zooniverse project(s) & 13\% & 	38 \\
\end{tabular}
\caption{\label{tab:more}
Response to the question ``Have you subsequently participated in other citizen science projects?''
}
\end{table}

A significant minority (37\%) of respondents had browsed the Talk form, showing that while of interest to many, it was far from ubiquitous. 
The fact that so many did not refer to the forum suggests that the 
majority were able to perform the classification exercises 
without recourse to the additional information on those discussion boards.
The primary reason stated for posting to the boards was to discuss 
findings with other Citizen Scientists.

Most respondents reported that as a result of the project they were motivated to engage more fully with science
(\cref{tab:future}) and the majority also
went on to work with other citizen science projects (\cref{tab:more}).

Overall the project was found to have a very positive response from respondents, with most having benefited from their engagement, and an overwhelming majority (more than 97\%) were keen to continue participation in a future CERN physics project.

Further analysis of the citizen science click data will be performed by school children
in collaboration with the UK charity the Institute for Research in Schools (IRIS)~\cite{iris}. 
At the time of writing 61 schools had signed up for this project through IRIS.

\section{Conclusion}
\label{sec:conclusion}

The first mass participation citizen science project for the Large Hadron Collider has been extremely successful. 
More than \nvolunteers{} citizen scientists participated, with a wide range of 
ages, backgrounds, and geographical spread represented. More than 1.4 million features of interest were identified in images from the ATLAS detector.

A study of behaviour showed that most Citizen Scientists classified just a handful of images, 
though a minority classified hundreds or thousands. 
A dedicate discussion forum allowed Citizen Scientists to interact with one another,
and with the project scientists. 
The vocabulary used in the forum ranged from basic visual features to highly 
abstract and technical terms.
The frequencies of some words in particular contexts indicated a distinct
technical vocabularly emerging from the Citizen Scientists' discussions 
-- one which would not immediately be understood by professional scientists in the field.

The societal impact was evaluated from a dedicated survey, with a very positive response. 
Almost two thirds of respondents were motivated to find out more about science directly from the project, 
while 97\% of respondents would like to see a follow-up project with more CERN data. 

The classification data from the citizen scientists have been released 
for final analysis by school pupils, in collaboration with the Institute for Research in Schools.

\section*{Acknowledgements}




The \url{HiggsHunters.org} project is a collaboration between the University of Oxford and the University of Birmingham in the United Kingdom, 
and NYU in the United States.  It makes use of the \href{www.zooniverse.org}{Zooniverse} citizen science platform, which hosts over 40 projects from searches for new 
astrophysical objects in telescope surveys to following the habits of wildlife in the Serengeti. The HiggsHunters project shows collisions 
recorded by the ATLAS experiment and uses software and display tools developed by the ATLAS collaboration. 
The authors gratefully acknowledge the generous financial support of the UK Science and Technology Facilities Council, the University of Oxford, and Merton College, Oxford.
The project would have been impossible without the dedicated engagement
of the many HiggsHunters volunteers and in particular the moderators.
We are grateful to Pete Watkins for helpful comments and suggestions.


\bibliography{references,ATLAS,CMS}{}
\bibliographystyle{vancouver}

\end{document}